\documentclass[twocolumn]{aastex701}

\begin{document}

\title{Confirming the magnetic field detection at the surface of $\chi$ Cyg.}

\author[0000-0001-8477-5265]{Alexis Lavail}
\affiliation{Institut de Recherche en Astrophysique et Plan\'etologie, Universit\'e de Toulouse, CNRS, CNES, UMR 5277, 14 avenue \'Edouard Belin, 31400 Toulouse, France}
\email[show]{astro@lavail.net}  

\author[orcid=0000-0002-1097-8307]{Arturo L\'{o}pez Ariste} 
\affiliation{Institut de Recherche en Astrophysique et Plan\'etologie, Universit\'e de Toulouse, CNRS, CNES, UMR 5277, 14 avenue \'Edouard Belin, 31400 Toulouse, France}
\email{arturo.lopez-ariste@cnrs.fr}

\author[0000-0002-1257-6782]{Quentin Pilate}
\affiliation{Institut de Recherche en Astrophysique et Plan\'etologie, Universit\'e de Toulouse, CNRS, CNES, UMR 5277, 14 avenue \'Edouard Belin, 31400 Toulouse, France}
\email{quentin.pilate@utoulouse.fr}

\author[0000-0001-7748-0178]{Philippe Mathias}
\affiliation{Institut de Recherche en Astrophysique et Plan\'etologie, Universit\'e de Toulouse, CNRS, CNES, UMR 5277, 57 avenue d’Azereix, 65000 Tarbes, France}
\email{philippe.mathias@utoulouse.fr}

\author[0000-0003-2977-5072]{Fabrice Herpin}
\affiliation{Laboratoire d'astrophysique de Bordeaux, Univ. Bordeaux, CNRS, B18N, all\'ee Geoffroy Saint-Hilaire, 33615 Pessac, France}
\email{fabrice.herpin@u-bordeaux.fr}

\author[0000-0001-9600-8258]{Agn\`es L\`ebre}
\affiliation{Laboratoire Univers et Particules de Montpellier, Universit\'e de Montpellier, CNRS, 34095 Montpellier, France}
\email{agnes.lebre@umontpellier.fr}

\begin{abstract}
We present spectropolarimetric observations of $\chi$ Cygni obtained with Neo-Narval at T\'elescope Bernard Lyot in 2025. We obtained observations across three epochs (2025 Jul, Aug, and Oct) near maximum light to search for magnetic field signatures at the stellar photosphere. We detected a clear circular polarization signal in the 2025 Aug observations (pulsation phases $0.99$ to $0.01$). We measure a mean longitudinal magnetic field of $B_l = 3.4 \pm 0.6$ G. No detections were obtained for the 2025 Jul and Oct epochs. The pulsation-phase dependence suggests that field detection is tied to specific shock conditions near maximum light.  

\end{abstract}



\section{Introduction}

The star $\chi$ Cygni is a bright AGB variable with a pulsation period of $\sim$408 days \citep{Samus2017} and  a magnitude peak-to-peak amplitude range of 3.3--14.2 \citep{Samus2017}. This large pulsation amplitude is also present in the radial velocity curves, and is associated to strong radiative shocks. \citet{Lebre2014} reported the first detection of a magnetic field in a Mira star. They observed $\chi$ Cyg near its maximum light in 2012 (around pulsation phase $\phi = 0.96$) using the Narval optical spectropolarimeter mounted at \textit{T\'elescope Bernard Lyot} (TBL) in the French Pyrenees. Using the Least Squares Deconvolution (LSD) technique \citep{1997MNRAS.291..658D,2010A&A...524A...5K}, they detected a weak circular polarization (Stokes $V$) signature with an amplitude of $2 \times 10^{-5}$ of the intensity continuum. This signature, only present on the blue side of the blue component of the intensity line, corresponds to a longitudinal magnetic field of 2--3~G at the photospheric level, which they attributed to shock amplification of an underlying stellar magnetic field. 

This detection is the only detection of a photospheric magnetic field in a Mira star to date, despite a decade of monitoring of around ten Mira stars at TBL.  In stark contrast, magnetic fields in the circumstellar envelopes of AGB stars are routinely detected through polarization observations of radio masers. SiO maser polarimetry has revealed magnetic fields of several Gauss in the inner circumstellar envelope at a few stellar radii \citep{Herpin2006,2024A&A...688A.143M}, while H$_2$O and OH masers trace fields extending to hundreds and thousands of stellar radii, respectively \citep{Vlemmings2006}. This discrepancy between the scarcity of spectropolarimetric detections at the photospheric level and the relatively large number of maser-derived magnetic fields challenges our understanding of AGB stars' magnetism.

Here we report new spectropolarimetric observations of $\chi$ Cyg obtained with Neo-Narval at TBL in 2025. Following the same data analysis process as described by \citep{Lebre2014}, we also detect a magnetic field at the surface of $\chi$ Cyg near maximum light with similar amplitude. 

\section{Data and methods}
Our observations were acquired in three epochs in 2025 Jul, Aug, and Oct. For each epoch, we observed the star multiple times over several nights. Individual observations were taken with exposure times of $4\times19$ s. We obtained 177 exposures between 2025 Jul 17--18 (totalling 3.7 hours of integration time), 240 exposures between 2025 Aug 08--16 (5.1 hours), and 131 exposures between 2025 Oct 16--26 (2.8 hours).  

Neo-Narval is an echelle spectropolarimeter covering the optical range between 380 -- 1050 nm at $R \approx 65 000$. Details about Neo-Narval and its data reduction are laid out in \citet{2022A&A...661A..91L}. Reduced data are made publicly available after a 1-year proprietary period in the PolarBase\footnote{\url{https://www.polarbase.ovgso.fr/}} archive. The data are immediately available at this Zenodo record\footnote{\url{https://doi.org/10.5281/zenodo.17665025}} \citep{lavail_2025_17665025} with other scripts and data (pre-processing scripts, LSD line list, LSD profiles) to reproduce this work. 

\begin{figure*}[ht!]
\plotone{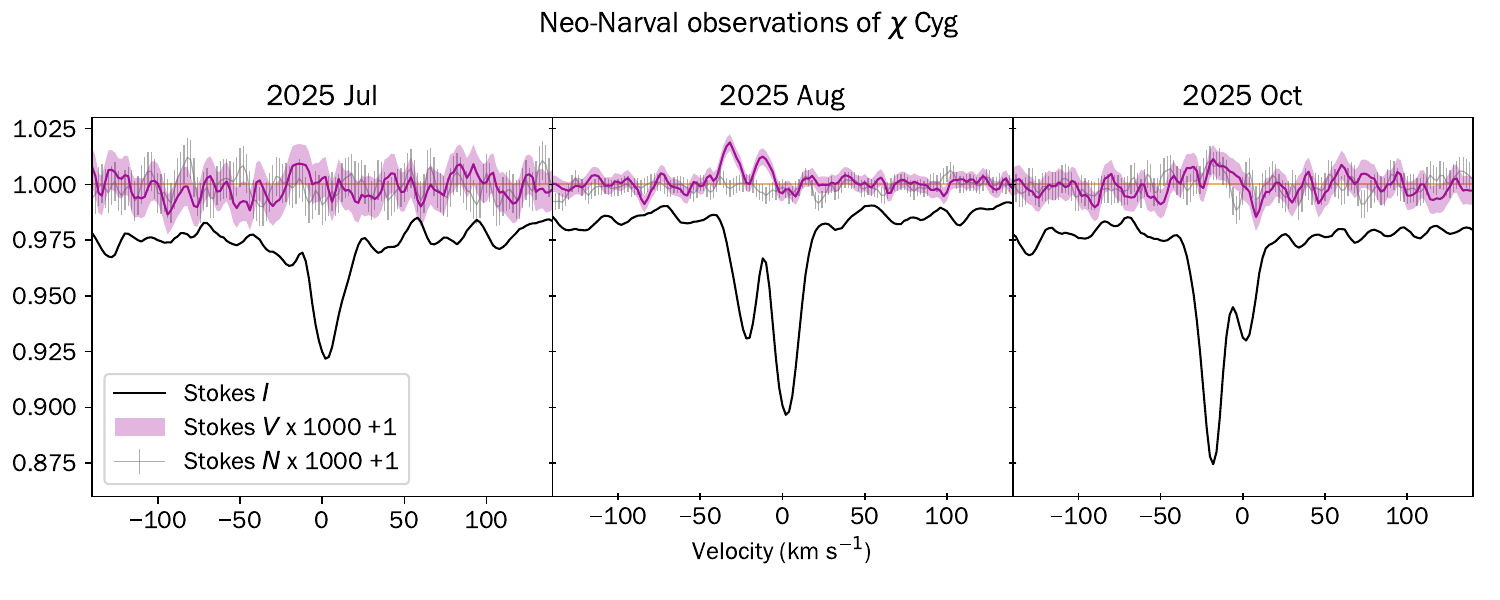}
\caption{Stokes $IVN$ LSD profiles of $\chi$ Cyg obtained with Neo-Narval in 2025. The error bars on Stokes $N$ and the shaded area on Stokes $V$ represent the 1-$\sigma$ confidence interval.}
\label{fig:stokes}
\end{figure*}

First, we co-added all individual spectra for each epoch to obtain very-high SNR Stokes $I$ (intensity), $V$ (circular polarization), and $N$ (null polarization) spectra. We then applied the Least Squares Deconvolution (LSD) technique \citep{1997MNRAS.291..658D,2010A&A...524A...5K} to the co-added spectra. LSD is a multiline technique that combines information from thousands of spectral lines to produce mean line profiles with significantly enhanced signal-to-noise ratio. The LSD profiles were computed using a line mask appropriate for a star with effective temperature of $3500~K$ and surface gravity of log($g$) = 0.5. We extracted the line list from VALD3 \citep{2015PhyS...90e4005R}, kept spectral lines deeper than 20\% of the continuum with know effective Land\'e factor ($g_{\textrm{eff}}$). We ended up with a list of 12447 spectral lines, with a mean wavelength of $4888.5$~\AA, mean $g_{\textrm{eff}}$ of 1.2, and mean depth of 0.59. Our LSD Stokes $IVN$ profiles for Jul, Aug, and Oct are shown in Fig.~\ref{fig:stokes}.  Our LSD implementation follows \citep{2010A&A...524A...5K}. 

\section{Results and conclusions}
We obtain a clear magnetic field detection for 2025 Aug, but no detection for Jul and Oct. Our Jul exposures were taken at pulsation phases $\phi$ 0.93 -- 0.94, Aug exposures at $\phi$ 0.99 -- 0.01, and the Oct exposures at $\phi$ 0.16 -- 0.19. 

We measure a mean longitudinal magnetic field $B_l = 3.4 \pm 0.6$ G slightly higher than the $2$--$3$ G reported by \citep{Lebre2014}. This second magnetic field detection at the surface of $\chi$ Cyg gives confidence that there is a magnetic field present at peak luminosity and can help up inform models of magnetic field formation and amplification on Mira stars, particularly in conjunction with shocks.

%% Please use the acknowledgment and contribution environments. This will 
%% be anonomyized when the "anonymous" style option is used. 
\begin{acknowledgements}
This work has made use of the VALD database, operated at Uppsala University, the Institute of Astronomy RAS in Moscow, and the University of Vienna.
\end{acknowledgements}

\begin{contribution}
%%This section gives authors the space to recognize author contributions. The text inside this environment is NOT counted towards the total word quanta. At a minimum, manuscripts are expected to include this text:

ALa wrote the observing proposal, led the data analysis and the writing of the manuscript.
ALAr worked on the instrumentation, data reduction, and participated in the data interpretation.
QPe wrote the observing proposal and participated in the data interpretation.
PMa planned the observations and participated in the data interpretation.
FHe coordinates joint radio observations and helped data interpretation.
AL\`e led the first study of $\chi$ Cyg, planned observations, and helped with data interpretation. 
All co-authors edited the manuscript.

\end{contribution}

\facilities{TBL(Neo-Narval)}

\software{
        NumPy \citep{harris2020array},
        Matplotlib \citep{Hunter:2007},
        astropy \citep{2013A&A...558A..33A,2018AJ....156..123A,2022ApJ...935..167A},  
        gnu-parallel \citep{10.5555/3235180}, 
        SciPy \citep{2020SciPy-NMeth}
          }

%% For this sample we use BibTeX plus aasjournalv7.bst to generate the

\bibliography{chicyg}{}
\bibliographystyle{aasjournalv7}



\end{document}